# Synthesis by Size Focusing of Lithium Tantalate Nanoparticles with a Tunable Second Harmonic Optical Activity

*Rana Faryad Ali[†], Byron D. Gates*[*,†]*

[†] Department of Chemistry and 4D LABS, Simon Fraser University, 8888 University Drive, Burnaby, BC, V5A 1S6, Canada

[*] E-mail: bgates@sfu.ca

**Abstract**

Nonlinear optics at the nanoscale has emerged as a sought-after platform for sensing and imaging applications. The development of these materials is having an impact on fields that include advanced information technology, signal processing circuits, and cryptography. Lithium tantalate ($LiTaO_3$) is an attractive nonlinear optical material due to its high optical damage threshold (e.g., tolerance to >500 MW/cm$^2$ from a nanosecond pulsed laser) and broad range of ultraviolet-visible (UV-Vis) transparency (i.e., 0.28–5.5 µm) relative to many other nonlinear optical materials (e.g., niobates, titanates). Despite many synthetic reports on metal oxides, very little is known about the preparation of uniform, crystalline $LiTaO_3$ nanoparticles (NPs) of a pure phase, as well as details on their mechanism of nucleation and growth. In this article, we introduce a solution-phase method for the preparation of $LiTaO_3$ NPs with tunable dimensions. This solution-phase process results in the formation of crystalline, uniform NPs of $LiTaO_3$ of a pure phase when carried out at 220 °C. This method can prepare crystalline $LiTaO_3$ NPs without the need for further heat treatment or the use of an inert atmosphere. Results presented herein also provide insights into the growth mechanism of these NPs. The reaction included the processes of oriented attachment and Ostwald ripening. The results of our study also indicate that the growth of the $LiTaO_3$ NPs was a result of a "size focusing" effect, which enables the ability to tune their diameters from 200 to 500 nm. The crystalline NPs were optically active towards second harmonic generation (SHG). These studies deepen our understanding of the methods by which NPs can be prepared from metal oxides. These studies specifically demonstrate the preparation of optically active $LiTaO_3$ NPs of uniform and controllable dimensions that could be used in a broad range of fundamental studies and applications in nanophotonics.

**Introduction**

This study demonstrates a solution-phase method to address the challenges to prepare uniform, crystalline lithium tantalate (LiTaO$_3$) nanoparticles (NPs) of a pure phase and with tunable dimensions. Lithium tantalate belongs to a group of oxides with the formula ABO$_3$ (A = Li; B = Ta) that have attracted attention due to their wide range of applications in the fields of ferroelectrics,[1,2] piezoelectrics,[3,4] photocatalysis,[5] photonics, and nonlinear optics (NLOs).[6] In NLOs, LiTaO$_3$ based materials have attracted attention over other materials (e.g., lithium niobate, lithium borate) due to their noticeably lower birefringence and higher resistance to optical damage. Because of their excellent NLO response, LiTaO$_3$ containing nanostructures are important components of waveguides,[7,8] wavelength conversion devices,[9] ultrafast lasers,[10] and holographics and other optical devices.[11] Recently, LiTaO$_3$ NPs have also been explored for use as second harmonic generation (SHG) imaging probes for their potential to expand SHG-based microscopy techniques. For example, these NPs have been used as probes to assist in imaging cells by taking advantage of the exceptional SHG properties of these particles.[12,13] The preparation of uniform, size-tunable, and single-crystalline LiTaO$_3$ NPs is necessary for many of the targeted applications. To meet the needs of applications that seek to utilize these properties, different synthetic routes have been sought to prepare uniform LiTaO$_3$ NPs.

Many solid-state approaches, molten salt syntheses, and sol-gel methods have been reported for the preparation of LiTaO$_3$ based materials.[14–21] These methods have been limited in their ability to tune the size of the LiTaO$_3$ NPs and to overcome aggregation in the products and often require high-temperature treatment (>500 °C) that results in the inclusion of impurities in the products. Often these products include micrometer-sized particles or compositional impurities. In recent years, a few solution-phase approaches have been developed for the preparation of LiTaO$_3$

NPs that have been able to achieve minimal aggregation of the products.[22–29] These solution-phase approaches include some relatively low-temperature methods for the preparation of crystalline particles with dimensions below 100 nm. None of these solution-phase methods are, however, able to provide a good control over the size, uniformity, shape, and purity of the $LiTaO_3$ nanomaterials.

The opportunities and challenges described above motivate us to develop a method capable of preparing $LiTaO_3$ NPs with a finer control over their size, uniformity, and purity while also producing a product with minimal aggregation. We sought to develop an alternative synthetic method through solvothermal assistance due to its widespread use in materials synthesis and the availability of the necessary metal alkoxide precursors that are soluble in non-aqueous solvents (e.g., benzyl alcohol used in this study). While several solvothermal methods are available to prepare NPs in the field of materials science, a limited number of reports are available to synthesize $LiTaO_3$ NPs. The previously reported solvothermal methods for the preparation of $LiTaO_3$ NPs have limitations that include the formation of aggregated products, the inclusion of impurities (e.g., tantalum oxide), relatively high reaction temperatures (e.g., 250 °C), and/or a lack of size tunability.[23,24,27,30] The solvothermal method reported herein sought to address these challenges and to provide a versatile route to prepare uniform, crystalline, and phase-pure $LiTaO_3$ NPs with tunable dimensions. The formation of NPs can be influenced by a number of factors that include the kinetics of the transformation of reagents, particle nucleation and growth, and potentially competing thermodynamically stable states. The role of reaction time and, thus, the effects of nucleation and growth, was systematically evaluated for its potential influence on the formation of the $LiTaO_3$ NPs.

The surfactant-assisted, solution-phase synthesis of $LiTaO_3$ NPs reported herein was found to proceed through a process of Ostwald ripening that can provide a size tunable preparation. This one-step solvothermal method to prepare LiTaO$_3$ NPs offers a number of advantages, such as processing times as short as 4 days (d) and reaction temperatures down to 220 °C without the need for further heat treatment or the use of an inert atmosphere. The LiTaO$_3$ NPs prepared by this route were assessed as a function of reaction time (e.g., 4 to 7 days) to assess the evolution of their dimensions and shapes. The size and crystallinity of the product, as well as its ability to form a dispersion upon a substrate were characterized by transmission electron microscopy (TEM) techniques. The composition, phase, and purity of the products were further characterized by X-ray diffraction (XRD) and Raman spectroscopy techniques. The results of these analyses indicated that the product contained crystalline LiTaO$_3$ NPs of a pure rhombohedral phase. In addition, the optical SHG response over visible wavelengths of these LiTaO$_3$ NPs were characterized by adjusting the wavelength of an incident pulsed laser.

**Results and Discussion**

We developed a surfactant-assisted solution-phase method to prepare LiTaO$_3$ NPs with uniform sizes and shapes. This solution-phase synthesis used tantalum ethoxide [Ta(OC$_2$H$_5$)$_5$] and lithium hydroxide monohydrate (LiOH·H$_2$O) as precursors. These reagents were dissolved in a solution of benzyl alcohol. Solvothermal treatment of the precursors was performed at 220 °C [i.e., above the boiling point of benzyl alcohol (205 °C)] over a period from 3 to 7 d. The selection of this polar, alcohol as a solvent was based on prior art that demonstrated its utility towards similar reactions, assisting in the dissolution of precursors, and an ability to regulate the reactivity of the precursors to enable control over the reaction rates.[31] The reaction mixture also contained triethylamine, which served as a surfactant during the formation of the LiTaO$_3$ NPs. This short-chain tertiary amine was selected to passivate the surfaces of the nanocrystals during their growth and to minimize aggregation of the resulting NPs. Triethylamine has also been demonstrated for its utility in regulating the growth of other types of NPs.[31]

A mechanism for the growth of the LiTaO$_3$ NPs is proposed in Figure 1 based on the empirical data collected in this study as a function of reaction time. The products prepared at 3 d and 4 d contained agglomerated NPs with average diameters <10 nm (Figures 2a, S1 and S2). Some larger NPs with diameters between ~200 and ~300 nm were also observed in the reaction product obtained at 4 d, indicating a bimodal size distribution at this stage of the reaction (Figure 2a). This result is distinct from what is observed during processes that are driven only by Ostwald ripening. This increase in dimensions of the NPs with prolonged reaction times up to 4 d was instead attributed to the agglomeration of smaller NPs, which was mediated by the presence of the triethylamine surfactant. Agglomeration of the smaller NPs is attributed to their relatively high chemical potential due to their relatively large surface-to-volume ratio.[32–35] These small NPs also have a greater mobility in solution, which increases their frequency of collision and can enhance their probability to agglomerate. A further increase in the reaction time to 5 d resulted in the formation of a relatively well-dispersed and uniform product of LiTaO$_3$ NPs with average dimensions of 200 ± 5 nm. This further increase in particle size and an improvement in the uniformity of the size distribution of the products were attributed to the further growth of NPs that proceeded through processes that included oriented attachment and Ostwald ripening. During Ostwald ripening, the higher chemical potential of the smaller NPs relative to the larger NPs leads to a faster dissolution of the smaller particles in the solution.[32–35] The larger NPs are more stable due to their smaller chemical potential, and they tend to progressively grow into larger NPs at the sacrifice of the smaller particles. The smaller NPs could dissolve into monomeric or similar species in the solution, and the larger NPs subsequently grow from addition of these highly reactive species onto their surfaces. Under normal conditions, a process dominated by Ostwald ripening produces a unimodal size distribution, which is distinct from the size distribution observed in the early stages of this reaction. The size distribution of the NPs grown by an Ostwald ripening process can also broaden and shift to larger dimensions during subsequent coarsening processes.[32–35] Multiple mechanisms for particle formation and growth appear to be taking place during this solvothermal synthesis of LiTaO$_3$ NPs.

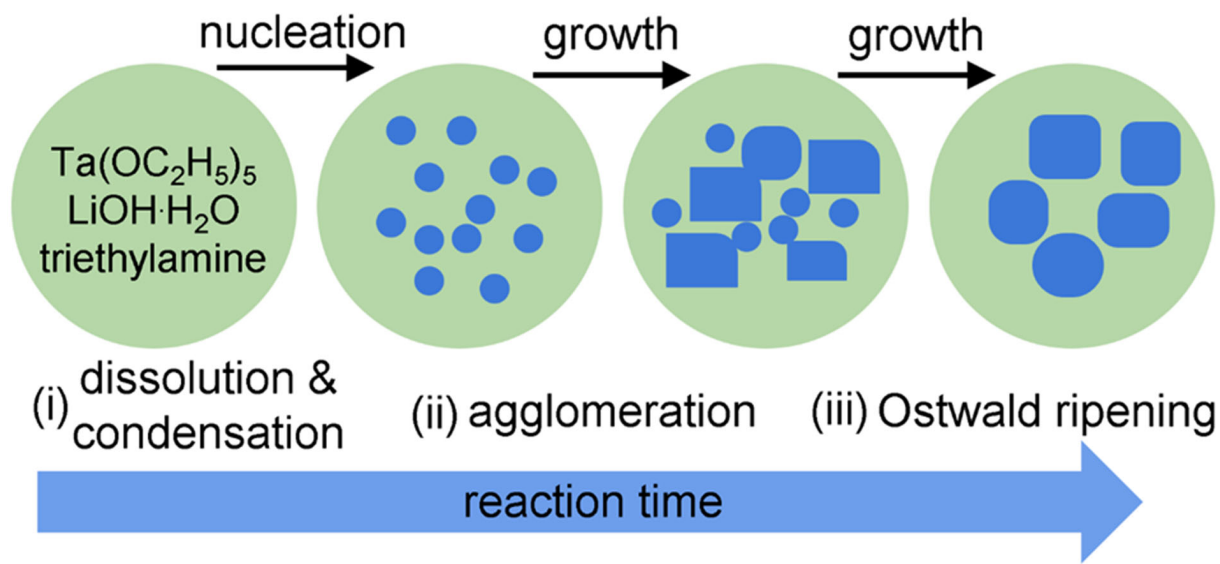

**Figure 1.** A proposed mechanism for the synthesis of lithium tantalate (LiTaO$_3$) nanoparticles (NPs) through a solvothermal process. The solvothermal treatment of the precursors produced nuclei that formed NPs of LiTaO$_3$. Growth of these NPs proceeded through particle agglomeration, but Ostwald ripening dominated their subsequent growth during a prolonged heat treatment.

The growth of LiTaO$_3$ NPs at the later stages of the reaction appears to proceed through the deposition of material either onto individual NPs or onto agglomerates of these NPs. The mechanism of growth was assessed by analyzing the structure, size and morphology of the resulting NPs as a function of the reaction time. Nanoparticles prepared through a process of oriented attachment followed by Ostwald ripening usually consist of a single crystalline domain.

Ostwald ripening of agglomerates (or aggregates) of NPs will, however, result in the formation of polycrystalline particles that can potentially exhibit twinned crystal planes. The NPs obtained at a reaction time of 5 d were polycrystalline, which supported the hypothesis that these NPs aged through a process that could include both oriented attachment and Ostwald ripening. During the aging process, polycrystallinity in the product can arise from twin boundaries and strained structures that form as a result of differences in the surface energy of faceted particles.[36–38] And even proceeding particle growth, during the nucleation process the proportion of colloidal seeds or nanoparticles with a bimodal size distribution may change through Ostwald ripening. The presence of solvents and ligands provides additional degrees of freedom and alters the diffusion to and from the surfaces of the seed particles. A one-pot solvothermal method, such as that explored herein, does not require the introduction of additional precursors (i.e., all the required precursors, solvents, and surfactants are included in one mixture) and may hinder the rate of atom deposition onto the seed particles. As a result, those seeds that grow faster (i.e., those exhibiting a relatively higher free energy) may reach a stable state earlier than their single-crystalline counterparts. The faster growing particles can, therefore, form a dominate product when the aging process proceeds through Ostwald ripening, but may also result in the formation of particles exhibiting stacking faults and twin boundaries.[36–38] When atom deposition is faster there may not be a significant difference in the size of the resulting seed particles. This scenario can lead to a preferential etching and ripening of defect-containing seeds and can favor the growth of single-crystalline seeds as a result of their differences in surface free energy.[36–38]

The bimodal size distribution observed at an early stage of the synthesis changed over the course of the reaction to a unimodal distribution (Figure 2). As the reaction time was extended from 3 and 4 days to 5 and 6 days, there was an apparent shift in the particle size distributions. For example, nearly all the small, seed-like particles were absent after 6 d. The variation in particle size within the samples also began to narrow after 4 d. The absolute standard deviation of the diameter of the products increased at longer reaction times, but the standard deviation relative to the product diameter decreased indicating a continual growth of the NPs and the formation of relatively uniform products at later stages of the reaction (Figure 3). Average dimensions of the $LiTaO_3$ NPs prepared at 6 d and 7 d were 300 ± 14 nm and 500 ± 20 nm, respectively. In this synthesis, the larger crystalline NPs appeared to grow at the expense of the smaller NPs present in the sample, which could result from the consumption of individual NPs or agglomerated species. As the growth proceeded at the elevated temperature of the solvothermal reaction, the smaller NPs contributed to further growth of the larger NPs. Although the average particle size increased as a function of reaction time, they retained their crystallinity. The driving force for this loss of the smaller particles, including those present in the agglomerates, was attributed to their higher chemical potential relative to that of the larger crystalline materials.[31]

In summary, the results suggest that the formation of $LiTaO_3$ NPs is initiated by a process of nucleation. As the reaction progressed, these relatively small nanocrystals start to agglomerate, which could be reversible as their surfaces remain passivated with the surfactant, triethylamine. Growth of individual NPs proceeds through a process of oriented attachment and Ostwald ripening to yield crystalline products. A secondary step of growth, following the initial agglomeration of the nuclei, transforms the product from a bimodal size distribution into a fairly uniform product at 5 d. The "size focusing" effect of Ostwald ripening also plays a pivotal role in preparing a uniform sample. The relatively high temperature and ambient atmosphere of the solvothermal synthesis could enable the Ostwald ripening process through dissolution of species from the surfaces of the nuclei or the dispersion and deposition of these smaller particles onto facets of the larger particles through oriented attachment. Ostwald ripening though is likely a dominate growth mechanism as the process of oriented attachment leads to the formation nanomaterials with distinct textures (e.g., nanorods, urchin-like structures). Narrowing of the size distribution from a broad range of sizes present in the initial product (e.g., at 4 d) through Ostwald ripening over the course of the reaction resulted in the formation of $LiTaO_3$ NPs with relatively uniform diameters. The sizes of the NPs could be adjusted as a function of reaction time. The thermodynamic driving force of the Oswald ripening process decreases as the size distribution narrows. As a result, this growth mechanism under the explored reaction conditions can be utilized to tune the size of the $LiTaO_3$ NPs from 200 to 500 nm.

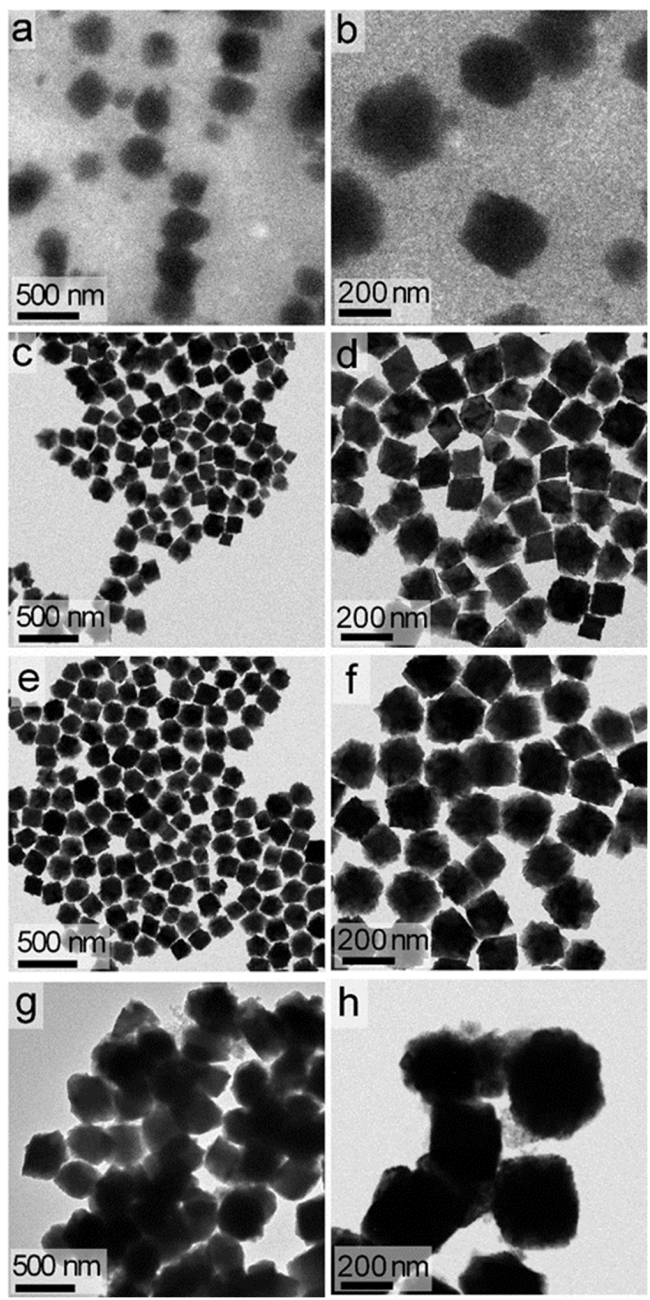

**Figure 2.** Transmission electron microscopy (TEM) analyses of LiTaO$_3$ NPs that were obtained after reaction times of: (a, b) 4 d; (c, d) 5 d; (e, f) 6 d; and (g, h) 7 d.

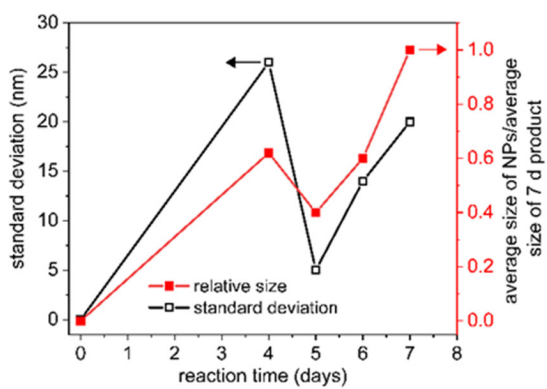

**Figure 3.** Red trace: particle size (normalized to the size of the particles obtained at t = 7 d) as a function of reaction time. Black trace: standard deviation of the particle sizes (in units of nanometers) as a function of reaction time.

Evolution of the phase and crystallinity of the NPs obtained at specific time points throughout the reaction were characterized using powder XRD analyses (Figure 4). The XRD patterns of the products obtained between 4 and 7 d indicated the formation of a crystalline product. All peaks were indexed to the formation of rhombohedral $LiTaO_3$ according to the ICSD reference material No. 9537.[39,40] Diffraction peaks were absent for the materials prepared at a reaction time of 3 d, which indicated an "amorphous nature" of the product or a very short range order in the particles obtained at the early stages of the reaction. The peak area of the (012) reflection for the nanoparticles increased faster as a function of reaction time relative to the peak areas for the other reflections. This result could indicate a more dominant growth of the NPs along this direction during the solvothermal treatment. The peak areas of the (104), (110), (024), and (116) reflections were smaller than the (012) reflection in the products, but these peak area ratios were larger than those observed for the reference standard (Table S1). The relative reflections observed for the NPs formed at early stages of the reaction indicated a less uniform growth of these products along the different crystalline planes. The ratio of the areas of the (104)/(012), (110)/(012), (024)/(012) and (116)/(012) peaks continued to change for the $LiTaO_3$ NPs after 5 d such that it indicated an enrichment of the {104}, {110}, {024} and {116} facets in the LiTaO$_3$ NPs with prolonged reaction times (Table S1). Longer reaction times resulted in a continual growth of the NPs along all facets, but these results also indicated a differential growth of the crystalline facets to that observed in the formation of bulk LiTaO$_3$ crystals. These observations further support the findings that the processes of oriented attachment and, importantly, Ostwald ripening each influenced the formation of the LiTaO$_3$ NPs during the solvothermal synthesis.

High-resolution TEM techniques were used to further assess the crystallinity of the sample obtained after 4 d of solvothermal treatment of the precursors. An analysis of the lattice fringes within this product was inconclusive, which was attributed to the high degree of disorder within these samples (Figure S2). Further analysis of the same product by SAED did, however, verify that the product obtained at 4 d was polycrystalline. Together with the outcome of the XRD analyses, the results suggest the formation of very small particles or nuclei at 4 d that contain short range order. The relative intensities of the (214), (024), and (116) reflections changed to a greater extent than the relative intensities of the (104) and (110) reflections for the products prepared at 5 d, which indicated a non-uniform growth of the NPs along the different crystalline planes. Longer reaction times resulted in an increase in the uniformity of the size distribution of the product as a result of further changes in the relative growth of the different crystalline facets (Table S1). High-resolution TEM and SAED analyses of the products collected between reaction times of 5 d and 6 d demonstrated the presence of regular lattice fringes and distinct diffraction patterns associated with these NPs (Figures S3, S4, S5 and S6). The dominant lattice spacings observed in these samples matched with the dominant reflections observed in the XRD diffraction analyses of these products. Regular lattice structures were observed throughout each of these NPs, which suggested the formation of single-crystalline products at the later stages of the reaction. A predominant periodic fringe pattern observed by HRTEM in these nanocrystals had a spacing of 3.7 Å. This spacing matched the inter-planar spacing for the (012) planes of LiTaO$_3$. Electron diffraction patterns obtained from individual LiTaO$_3$ NPs further indicated the single-crystalline nature of the product at reaction times between 5 d and 6 d (Figures S4 and S6). The analyses collectively indicated the presence of {012}, {104}, {116} and {110} as major facets within the individual single-crystalline nanoparticles of LiTaO$_3$ at the early stages of the reaction. Growth of each of the different crystal directions was more uniform at the later stages of the reaction, but the {012} facet remained the dominant growth direction. These analyses suggested that particle growth initiated and carried out through this solvothermal process was sufficiently controlled by the addition of the triethylamine as a surfactant to produce crystalline rhombohedral LiTaO$_3$ in the form of NPs with uniform dimensions.

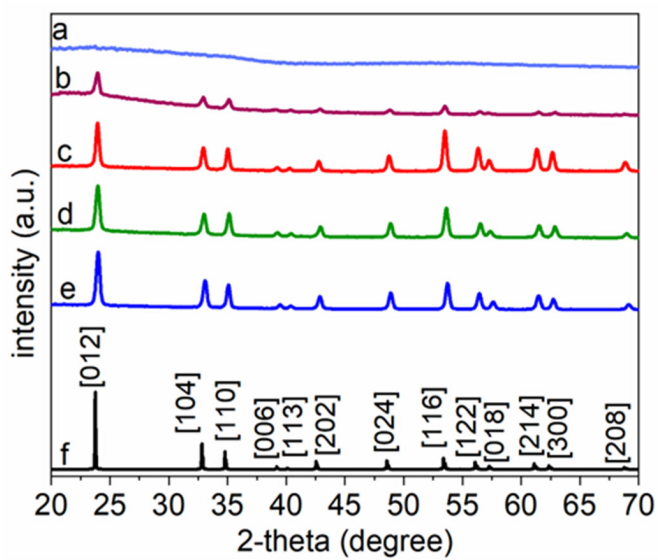

**Figure 4.** Powder X-ray diffraction (XRD) patterns of LiTaO$_3$ NPs obtained at reaction times of: (a) 3 d; (b) 4 d; (c) 5 d; (d) 6 d; and (e) 7 d. (f) The XRD pattern reported for rhombohedral LiTaO$_3$ (ICSD No. 9537) is included for reference.

The evolution of the phase and crystallinity of the LiTaO$_3$ NPs were evaluated further by Raman spectroscopy techniques (Figure 5). Characteristic peaks for LiTaO$_3$ were observed in the Raman spectra for samples prepared at reaction times from 4 d to 7 d, which were indexed to the formation of rhombohedral LiTaO$_3$. The Raman spectrum of the product prepared at a reaction time of 3 d indicated that the sample did not contain crystalline LiTaO$_3$ of sufficiently large domains for detection using our Raman spectrometer, and/or that the sample contained a relatively high degree of disorder at this early stage of the reaction. Raman spectroscopy also confirmed the purity of the LiTaO$_3$ NPs. This technique can be used to determine the phase and composition of the lithium tantalate product (e.g., LiTaO$_3$, Li$_5$TaO$_5$, Li$_3$TaO$_4$, and LiTa$_3$O$_8$).[41–43] The products of this solvothermal synthesis were not treated by additional high temperature processes (e.g., calcination), but instead were directly analyzed after the synthesis using only a purification step to isolate the products from any soluble by-products and excess surfactants as outlined in the Experimental Section. The Raman spectra for the LiTaO$_3$ NPs indicated the formation of a pure phase of rhombohedral LiTaO$_3$ after a reaction time of 4 d. The TEM, XRD, and Raman results collectively confirmed the formation of ~200-nm diameter crystalline, rhombohedral LiTaO$_3$ particles at 5 d and demonstrated that longer reaction times also produced crystalline rhombohedral LiTaO$_3$ particles with diameters up to ~500 nm.

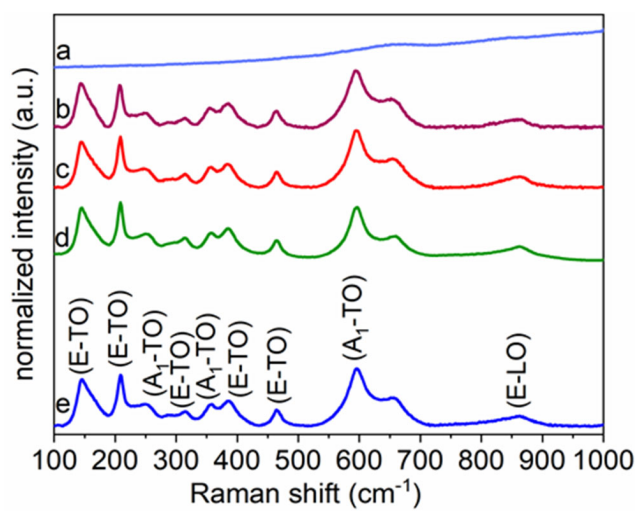

**Figure 5.** Analyses by Raman spectroscopy of the LiTaO$_3$ NPs obtained after a reaction time of: (a) 3 d; (b) 4 d; (c) 5 d; (d) 6 d; and (e) 7 d.

The optical second harmonic response of the LiTaO$_3$ NPs was also examined to evaluate their nonlinear optical (NLO) properties. Nonlinear optical materials have a wide range of important applications that include quantum light sources, frequency convertors, ultrafast optical switches, and memory storage devices. Recently, nanometer sized materials have received interest for their potential use as probes in SHG-based microscopy, in part due to their absence of "blinking" and their relatively consistent response across multiple frequencies, which are contrary to the properties of quantum dots.[44–46] Non-linear optical materials have also been sought for their ability to generate coherent light emission and for their stability to photobleaching relative to fluorescent probes (e.g., organic fluorophores and quantum dots).[47,48] Nanoscale NLO crystals can also provide benefits in the preparation of hybrid materials, such as through embedding these nanocrystals into easily processable polymers.[49,50] Another advantage of nanoscale NLO materials is their ability to generate a second harmonic response without requiring phase-matching conditions.[49,50] The intensity of the SHG in bulk NLO materials (e.g., where size of the crystals exceeds the wavelength of the incident light) is only significant if it obeys phase-matching conditions, which occurs if the second harmonic response moves in the material at the same velocity and in the same direction as the incident photons, resulting in a constructive interference. Nanoscale NLO crystals do not require this phase-matching condition for producing a significant second harmonic response due to their small diameters relative to the wavelength of the incident photons. As a result of not requiring phase-matching conditions, the SHG produced anywhere in the nanocrystal will be approximately in phase, allowing scattering to efficiently proceed regardless of the orientation of the nanoparticle.

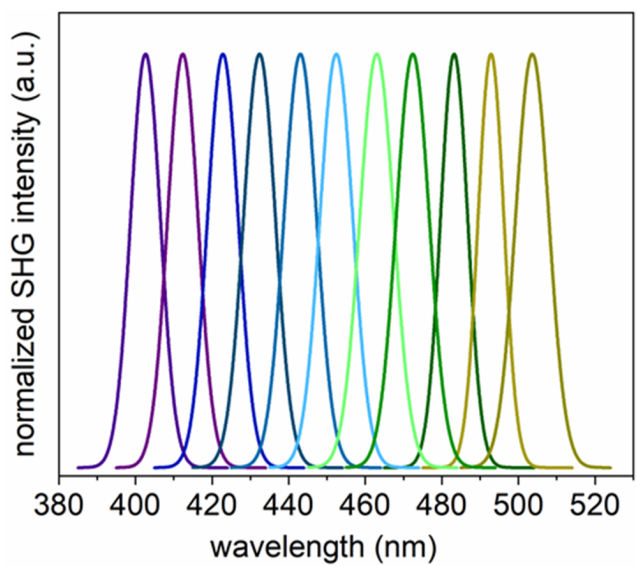

**Figure 6.** Spectra of second harmonic generation (SHG) obtained by pulsed laser irradiation of powdered samples prepared from ~200-nm diameter $LiTaO_3$ NPs supported on a glass coverslip. The results are plotted in false-colors corresponding to the tunable emission for the $LiTaO_3$ NPs with a second harmonic response centered at 400, 410, 420, 430, 440, 450, 460, 470, 480, 490 and 500 nm when excited at the corresponding fundamental wavelengths between 800 and 1,000 nm.

We selected the crystalline $LiTaO_3$ NPs prepared at 5 d by the solvothermal method for evaluating their nonlinear optical (NLO) properties. The NPs were evaluated for their SHG activity and the tunability of their SHG response using a series of different fundamental wavelengths (FW) as the incident light from a tunable femtosecond (fs) pulsed laser. The incident light was generated using a mode-locked Ti:sapphire laser with a pulse width of ∼140 fs and a tunable output of FWs from 680 to 1080 nm. The repetition rate and tuning speed of the fs pulses were 80 MHz and >40 nm s$^{-1}$, respectively. The SHG response for a powdered sample of the $LiTaO_3$ NPs was assessed for a series of discrete FWs while maintaining a constant incident power. The measured data points were normalized relative to their maximum intensity and fit using a type of Gaussian fit (i.e., spectroscopy function) in Origin Pro to estimate the peak position of each of the measured SHG responses.[51,52] A second harmonic response was observed at 400, 410, 420, 430, 440, 450, 460, 470, 480, 490 and 500 nm when the $LiTaO_3$ product was excited using FWs of 800, 820, 840, 860,

880, 900, 920, 940, 960, 980, and 1,000 nm, respectively. The results correlated well to the anticipated frequency doubling of the FWs (Figure 6). These SHG active NPs of LiTaO$_3$ can generate a second harmonic response over a broad range of wavelengths by adjusting the wavelength of the incident, fs pulsed laser. The SHG response of these NPs could potentially be extended to near-infrared and far-infrared wavelengths by choosing the appropriate wavelength for the incident laser. Another benefit of developing SHG materials, such as probes for use with SHG microscopy techniques, is that the full width at half maximum (FWHM) of the measured second harmonic response should correspond to $1/\sqrt{2}$ of the bandwidth of the fs pulsed laser.[53] An average FWHM of 7 nm was measured for the SHG from the LiTaO$_3$ NPs, which was significantly narrower than the FWHM of the emission signal reported for many fluorescent probes. Fluorescence markers, including quantum dots and organic dyes, have an emission with a spectral bandwidth (e.g., FWHM) of 20 nm to 50 nm.[54,55] The SHG active LiTaO$_3$ NPs could be used as imaging probes and could open up new opportunities in SHG microscopy techniques. It is anticipated that the "size-focusing effect" observed in the growth of LiTaO$_3$ NPs by a solvothermal process enabled with the inclusion of appropriate reagents including the selected solvent(s) and surfactant(s) can be extended to designing the syntheses of similar metal oxides systems (e.g., titanates, niobates, tungstates).

**Conclusion**

In summary, we demonstrated a surfactant assisted solution-phase method to prepare uniform, single-crystalline LiTaO$_3$ NPs with an average size between 200 nm and 500 nm, which is determined as a function of reaction time. The solution-phase process used to prepare these NPs was carried out at a relatively low temperature (e.g., 220 °C) in benzyl alcohol. During this solvothermal process, the precursors reacted to produce nuclei, which subsequently formed NPs of LiTaO$_3$. The growth of these NPs initially proceeded through a surfactant-controlled agglomeration of smaller particles. The processes of Ostwald ripening and oriented attachment likely played an important role in this reaction to control growth of the NPs while also creating a single-crystalline product. Uniform LiTaO$_3$ nanocrystals with a diameter of ~200 nm were prepared after 5 d of this reaction process. Their size could be tuned by increasing the reaction time. Due to the "size focusing" effect of the Ostwald ripening, the LiTaO$_3$ NPs maintained a relatively narrow size distribution even after a reaction time of multiple days. Phase and purity of the products were characterized by XRD and Raman spectroscopy, which indicated the formation of LiTaO$_3$ of a pure rhombohedral phase at relatively low temperature reaction conditions. Although the LiTaO$_3$ NPs had a dominant growth along the (012) plane, a number of other crystal planes were equally preferred over the course of the reaction leading to the formation of multifaceted twinned crystalline particles. The resulting NPs were SHG active, which could be explored in the future for use as SHG imaging probes or contrast agents for applications and systems that require long-term monitoring (e.g., biological systems). The results observed in the formation of these LiTaO$_3$ NPs could provide insights to control the preparation of other types of NPs such as through offering guidance for the rational synthesis of other metal oxide nanocrystals that also have well-controlled compositions, dimensions, and crystallinity.

**Experimental Section**

All the chemicals were of an analytical grade and were used as received without further purification. Lithium tantalate NPs were prepared in a single-step solvothermal process. In brief, 40 mM tantalum ethoxide [(Ta(OC$_2$H$_5$)$_5$, >90%, Gelest Inc.] was dissolved in 10.0 mL of benzyl alcohol (99%, Acros Organics) and stirred for 30 min, which resulted in the formation of a pale yellow solution. This step was followed by the addition of 0.1 mL (i.e., 72 mM) of triethylamine [N(C$_2$H$_5$)$_3$, 99%, Anachemia], which served as a surfactant to assist in controlling the growth and colloidal stability of the nanocrystals. The mixture was stirred for another 30 min. After this period, 40 mM lithium hydroxide monohydrate (LiOH·H$_2$O, 99%, Alfa Aesar) was added to the solution, which was stirred for another 10 h at room temperature. The resulting mixture was transferred to a 23 mL Teflon lined autoclave (Model No. 4749, Parr Instruments Co., Moline, IL, USA) and heated at 220 °C for a specific period ranging from 3 to 7 d. After cooling to room temperature, white precipitates were isolated from the solution via a process of centrifugation (Model No. AccuSpin 400, Fisher Scientific) at 8000 rpm for 20 min and decanting of the solution. These solids were washed one time by resuspending with 10 mL of chloroform and two subsequent times with ethanol by repeating the processes of centrifugation and decanting of the solution. The product obtained at 3 d of solvothermal treatment was only washed with ethanol and water since the chloroform-based wash led to the complete dissolution of the samples. The purification process was repeated two more times with 10 mL of deionized water (18 MΩ·cm, produced using a Barnstead NANOpure DIamond water filtration system). The purified product was dried at 70 °C for 12 h to remove residual water prior to further analyses.

**Characterization of Lithium Tantalate NPs**

The morphology, dimensions, crystallinity, and lattice parameters of the LiTaO$_3$ NPs were characterized using an FEI Osiris X-FEG 8 transmission electron microscope (TEM) operated at an accelerating voltage of 200 kV. The TEM was calibrated using a thin film of aluminum before acquiring selected area electron diffraction (SAED) patterns from the samples. The camera length was 220 mm. Samples for TEM analyses were prepared by dispersing the purified products in ethanol followed by drop casting 5 μL of each suspension onto separate TEM grids (300 mesh copper grids coated with Formvar/carbon) purchased from Cedarlane Laboratories. Each TEM grid was dried at ~230 Torr for at least 20 min prior to analysis. The TEM apertures used to acquire SAED patterns from multiple NPs and the diffraction from a single nanoparticle were 40 μm and 10 μm, respectively.

Phase and crystallinity of the samples were further determined from X-ray diffraction (XRD) patterns acquired with a Rigaku R-Axis Rapid diffractometer equipped with a 3 kW sealed tube copper source (Kα radiation, λ = 0.15418 nm) collimated to 0.5 mm. Powder samples were packed into a cylindrical recess drilled into a glass microscope slide (Leica 1 mm Surgipath Snowcoat X-tra Micro Slides) for acquiring XRD patterns of the products.

Purity and phase of the product with respect to the desired rhombohedral phase were further assessed using Raman spectroscopy techniques. Raman spectra were collected using a Renishaw inVia Raman microscope with a 50× short-working-distance (SWD) objective lens and a 514 nm laser (argon ion laser, Model No. Stellar-Pro 514/50) set to 100% laser power with an exposure time of 30 s. The Raman spectrometer was calibrated by collecting the Raman spectrum of a polished silicon (Si) standard with a distinct peak centered at 520 cm$^{-1}$. The Raman spectra for the samples were acquired from 100 to 1,000 cm$^{-1}$ using a grating of 1,800 lines/mm and a scan rate of 10 cm$^{-1}$/s.

The SHG activity of the LiTaO$_3$ NPs were assessed using a Leica SP5 laser scanning confocal two photon microscope equipped with a Coherent Chameleon Vision II laser and a 63× objective lens (Leica, 1.0 NA). Dried powders of LiTaO$_3$ NPs were loaded onto glass coverslips and brought into the focal point of the microscope. The excitation wavelength was set between 800 and 1,000 nm, and the corresponding band-pass filters were centered between 400 and 500 nm, respectively, to selectively collect the SHG signals.

**Acknowledgements**

This work was supported in part by the Natural Sciences and Engineering Research Council of Canada (Discovery Grant No. RGPIN-2020-06522), and CMC Microsystems (MNT Grant No. 6345). This work made use of 4D LABS (www.4dlabs.com) and the Center for Soft Materials shared facilities at Simon Fraser University (SFU) supported by the Canada Foundation for Innovation (CFI), British Columbia Knowledge Development Fund (BCKDF), Western Economic Diversification Canada, and SFU.

**Conflict of Interest**

The authors declare no competing financial interest.

**Table of Contents (TOC)**

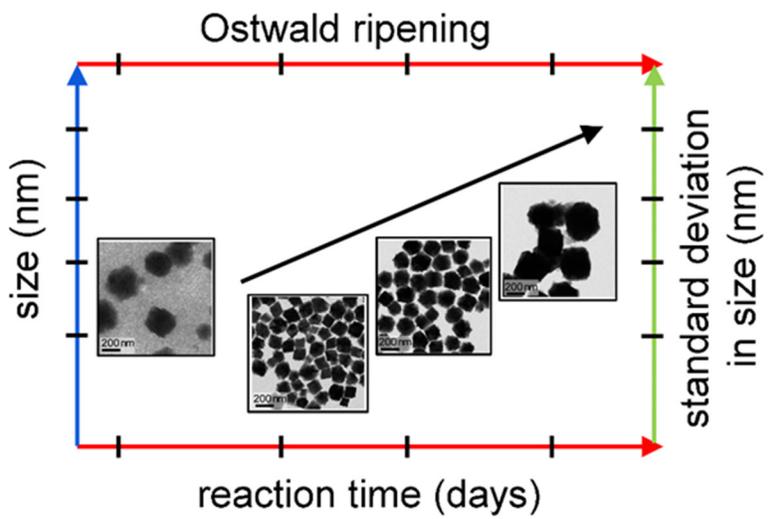

# Supporting Information

# Synthesis by Size Focusing of Lithium Tantalate Nanoparticles with a Tunable Second Harmonic Optical Activity

*Rana Faryad Ali[†], Byron D. Gates*[*,†]*

[†] Department of Chemistry and 4D LABS, Simon Fraser University, 8888 University Drive, Burnaby, BC, V5A 1S6, Canada

[*]  E-mail: bgates@sfu.ca

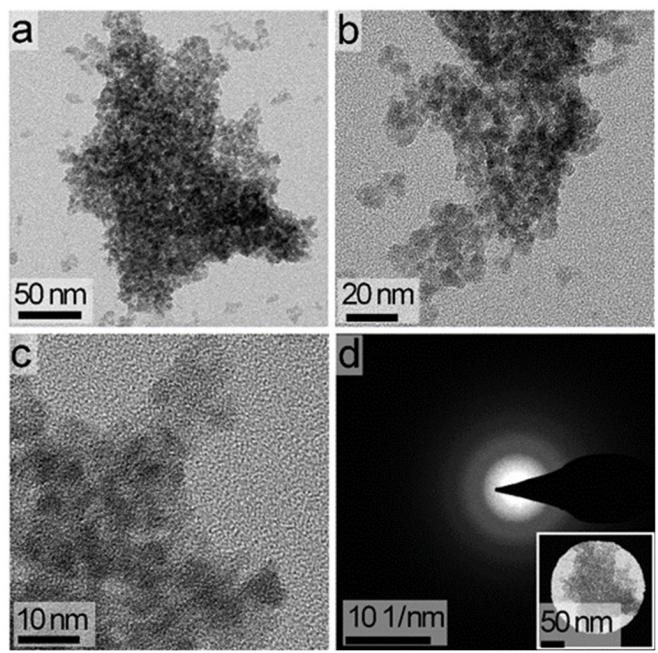

**Figure S1.** The product obtained after 3 d of solvothermal treatment of the precursors as characterized by: (a, b) transmission electron microscopy (TEM); (c) a high-resolution (HR) TEM (or HRTEM) analysis; and (d) a selected area electron diffraction (SAED) based analysis.

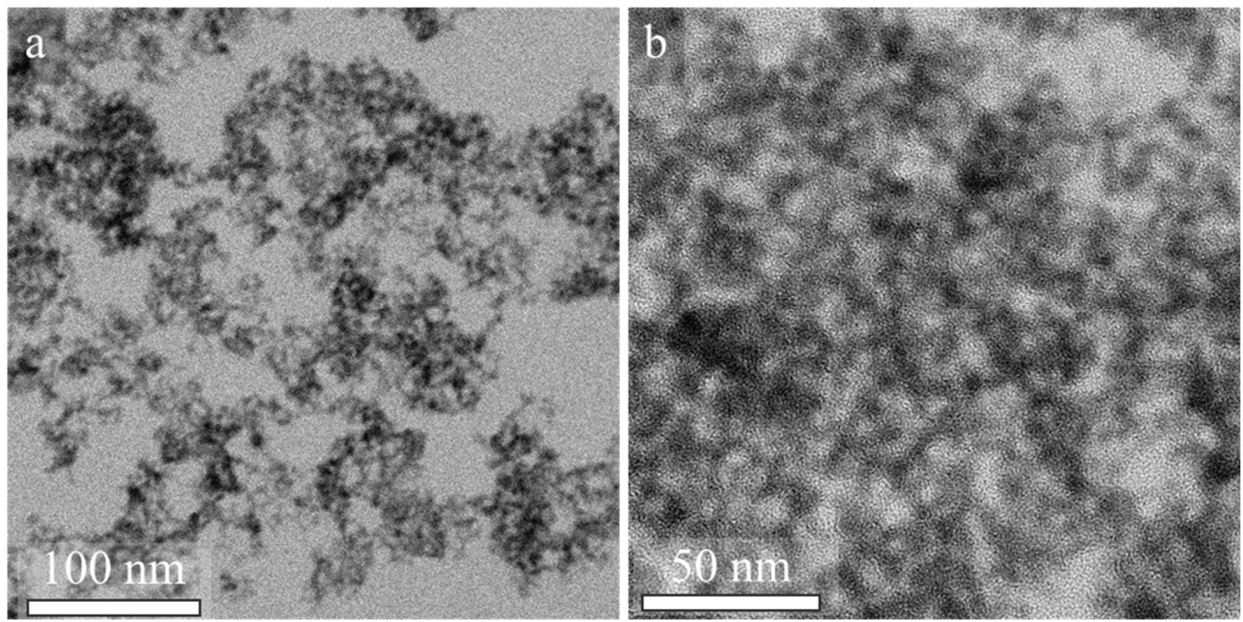

**Figure S2**. A HRTEM analysis of lithium tantalate (LiTaO$_3$) nanoparticles (NPs) obtained after a solvothermal treatment for 4 d, which indicates the formation of relatively small particles.

**Table S1.** Ratios the peak areas as measured by X-ray diffraction (XRD) relative to the (012) reflection for the reported $LiTaO_3$ reference material (ICSD No. 9537) and for NPs of $LiTaO_3$ prepared between reaction times of 4 d and 7 d.

| XRD peak ratios | Reference ICSD No. 9537 | 4 d product | 5 d product | 6 d product | 7 d product |
|---|---|---|---|---|---|
| (104)/(012) | 0.40 | 0.73 | 0.79 | 0.74 | 0.56 |
| (110)/(012) | 0.29 | 0.71 | 0.76 | 0.72 | 0.51 |
| (202)/(012) | 0.14 | 0.39 | 0.65 | 0.61 | 0.44 |
| (024)/(012) | 0.18 | 0.50 | 1.55 | 0.63 | 0.46 |
| (116)/(012) | 0.26 | 0.61 | 1.03 | 0.75 | 0.52 |
| (214)/(012) | 0.15 | 0.48 | 0.78 | 0.62 | 0.45 |

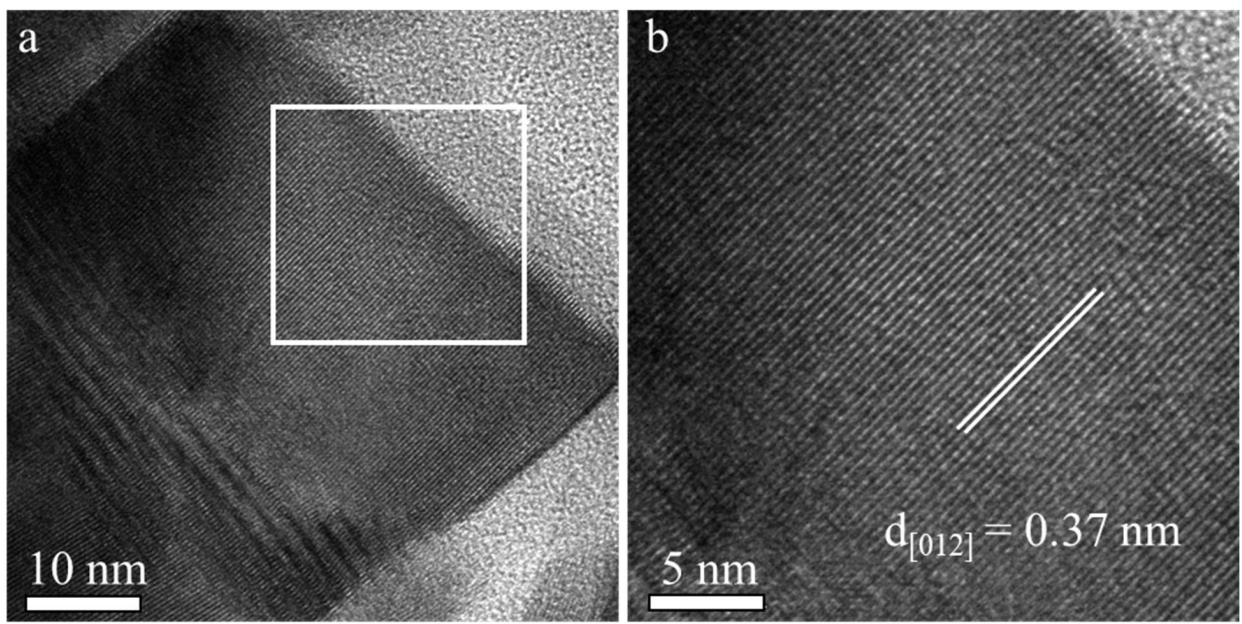

**Figure S3**. (a) Representative results from a HRTEM analysis of the LiTaO$_3$ NPs obtained at a solvothermal reaction time of 5 d. (b) A higher magnification of the sample obtained from the region indicated by the white box in (a). The white lines highlight the observed lattice fringe patterns, whose d-spacing is assigned on the image.

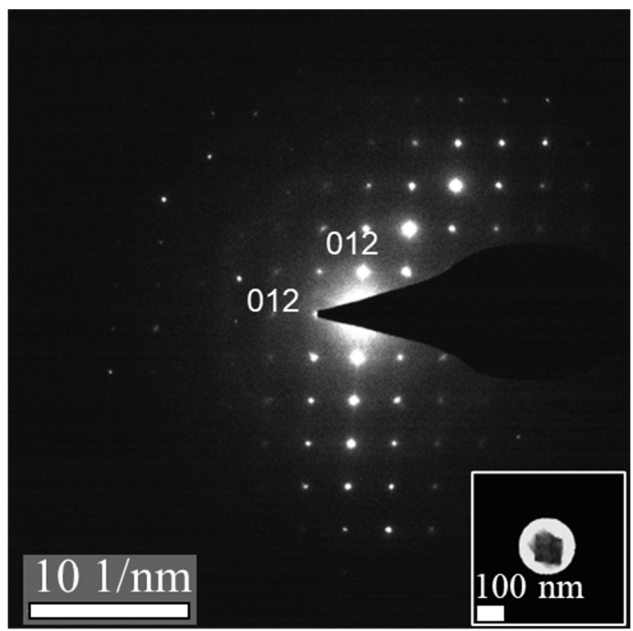

**Figure S4.** The selected area electron diffraction (SAED) pattern obtained from a single NP (shown within the inset). The SAED pattern indicates the crystalline nature of this LiTaO$_3$ NP, which was obtained after 5 d of solvothermal treatment.

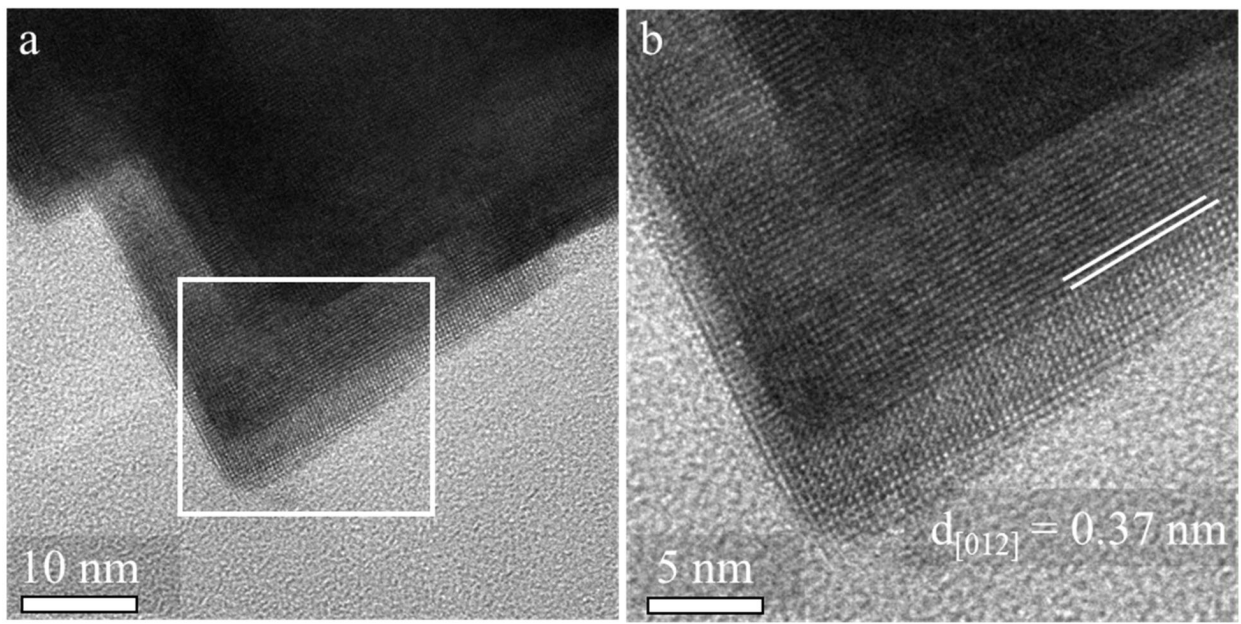

**Figure S5**. (a) Representative results from a HRTEM analysis of LiTaO$_3$ NPs obtained after a reaction time of 6 d. (b) A higher magnification of the sample obtained from the region indicated by the white box in (a). The white lines highlight the observed fringe patterns, whose d-spacing is assigned on the image.

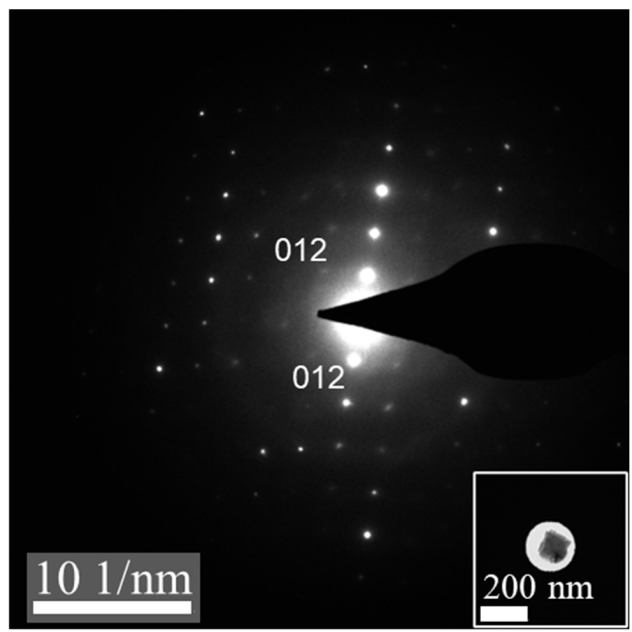

**Figure S6.** The selected area electron diffraction (SAED) from a single NP (shown within the inset). The SAED pattern indicates the crystalline nature of the LiTaO$_3$ NPs obtained after a reaction time of 6 d as discussed in further detail in the main text.